\documentclass[aps,prc,floatfix,twocolumn,nofootinbib,superscriptaddress]{revtex4-1}

\usepackage[]{graphicx}
\usepackage{color}
\usepackage{subfigure}
\usepackage{amsmath,amssymb,amsfonts}
\usepackage{tabularx}
\usepackage{float}
\usepackage{graphicx}

\usepackage{slashed}

\begin{document}

\title{Light quark energy loss in the flavor-dependent systems from holography}

\author{Le Zhang}
\affiliation{College of Physics and Electronic Science, Hubei Normal University, Huangshi, 435002, China}
\affiliation{Hubei Engineering Research Center for Micronano Optoelectronic Devices and Integration, Huangshi, 435002, China}
\affiliation{Hubei Key Laboratory of Optoelectronic Conversion Materials and Devices, Huangshi, 435002, China}

\author{Lu Yin}
\affiliation{College of Physics and Electronic Science, Hubei Normal University, Huangshi, 435002, China}

\author{Guo-Dong Zhou}
\affiliation{College of Physics and Electronic Science, Hubei Normal University, Huangshi, 435002, China}

\author{Chao-Jie Fan}
\affiliation{College of Physics and Electronic Science, Hubei Normal University, Huangshi, 435002, China}
\affiliation{Hubei Engineering Research Center for Micronano Optoelectronic Devices and Integration, Huangshi, 435002, China}
\affiliation{Hubei Key Laboratory of Optoelectronic Conversion Materials and Devices, Huangshi, 435002, China}

\author{Xun Chen}
\email{chenxun@usc.edu.cn}
\affiliation{School of Nuclear Science and Technology, University of South China, Hengyang 421001, China}

\date{\today}

\begin{abstract}
Using the holographic model of finite end point momentum shooting string approach, we study the instantaneous energy loss of light quarks for the flavor-dependent systems with $N_f = 0$, $N_f = 2$, and $N_f = 2+1$ in the Einstein-Maxwell-dilaton (EMD) framework. In particular, we investigate for the first time the impact of the flavor content of the strongly coupled quark-gluon
plasma medium on the instantaneous energy loss of light quarks. It turns out that the instantaneous energy loss of light quarks is smallest for $N_f = 0$, and adding $u(d)$ quarks and $s$ quark in the system increases this energy loss. In addition, we found that the instantaneous energy loss in the strongly coupled plasma is minimal near the critical temperature, but it increases as the system moves away from the critical end point due to rising temperature or chemical potential.
\end{abstract}

\maketitle
%
\begin{widetext}

\section{Introduction}
One of the main purposes of the heavy-ion collisions at the Large Hadron Collider (LHC) and the Relativistic Heavy-Ion Collider (RHIC) is to explore the properties of the strongly interacting quark-gluon plasma (QGP), a new state of matter consisting of deconfined quarks and gluons produced through high-energy collisions \cite {BRAHMS:2004adc, Shuryak:2004cy, STAR:2005gfr, Gyulassy:2004zy}. Jet quenching \cite {Wang:1992qdg, Majumder:2010qh, Qin:2015srf, Blaizot:2015lma} provides a useful tool for studying such QGP matter: hard partonic jets, which are produced by early hard scatterings, may interact with the constituents of the medium via elastic and inelastic collisions and lose their energy during their passage through the highly excited QCD matter. However, it is currently observed that the QGP behaves as a nearly perfect liquid \cite {PHENIX:2004vcz, STAR:2005gfr, Gyulassy:2004zy} with surprisingly small viscous effects. For example, the shear viscosity to entropy density ratio $\eta/s$ in strongly coupled holographic non-Abelian plasmas is close to a conjectured minimum bound $1/4\pi$~\cite { Kovtun:2004de, Buchel:2003tz}, which is an order of magnitude below the result of calculations in transport models involving ordinary hadrons~\cite { Demir:2008tr} or in QCD at weak coupling~\cite{ Arnold:2000dr}. This makes perturbative QCD generally inadequate~\cite{Baier:1996kr,Eskola:2004cr}. Lattice QCD is the main nonperturbative tool for studying strongly interacting QCD physics but also of limited utility, e.g., lattice approaches are very difficult to investigate the real time nonequilibrium phenomena.
On the other hand, the AdS/CFT correspondence~\cite{Maldacena:1997re, Gubser:1998bc, Witten:1998qj}  is regarded as an alternative nonperturbative approach for dealing with strongly coupling gauge theories.

AdS/CFT, which establishes a correspondence between $\mathcal{N} = 4$ supersymmetric Yang-Mills theory and type IIB super string theory formulated on $AdS_5 \times S^5$ space, has provided significant insights into the dynamics of various theories that share qualitative features with QGP (see Refs.~\cite{Casalderrey-Solana:2011dxg, DeWolfe:2013cua} and
references therein for recent reviews with AdS/CFT applications) .
During the past two decades, AdS/CFT has been applied to theoretical and phenomenological studies of jet quenching in strongly coupled plasma.
There currently exist a few holographic models for describing the energy loss of heavy and light quarks in strongly coupled plasma: the drag force~\cite{Gubser:2006bz, Herzog:2006gh} and diffusion coefficient ~\cite{Casalderrey-Solana:2006fio, Gubser:2006qh} for describing the heavy quark energy loss, and the jet quenching parameter~\cite{Liu:2006ug, Liu:2006he}, falling string~\cite{Chesler:2008uy, Chesler:2008wd,Arnold:2010ir, Arnold:2011qi, Ficnar:2012np}, and shooting string~\cite{Ficnar:2013wba,Ficnar:2013qxa} for light quark energy loss.  Additionally, holographic models are also used to study jet shape modifications and jet evolution~\cite{Hatta:2008tx, Chesler:2008hg}.

The five-dimensional Einstein-Maxwell-Dilaton (EMD) model, as a phenomenological gravity dual for hot and dense QCD matter, was proposed in Ref.~\cite {DeWolfe:2010he}.
In the black hole background, the EMD model is mapped onto the thermodynamic properties of QGP in the boundary field theory through the Hawking temperature and chemical potential of the  $AdS_5$ black hole solutions.
It has opened up a fruitful research area in the holographic application to the strongly coupled QGP. Currently, this EMD model has been widely applied to describe various aspects, such as the QCD phase diagram \cite{DeWolfe:2010he, He:2013qq, Yang:2014bqa, Yang:2015aia, Dudal:2018ztm, Chen:2018vty, Chen:2020ath, Chen:2019rez, Knaute:2017opk, Grefa:2021qvt, Cai:2022omk, Li:2023mpv, Rougemont:2023gfz, Zhao:2023gur, Fu:2024wkn, Jokela:2024xgz, Chen:2024edy, Chen:2023yug}, heavy quark potential \cite{Zhou:2020ssi}, transport properties \cite{DeWolfe:2011ts, Rougemont:2017tlu, Grefa:2022sav}, and jet quenching \cite{Rougemont:2015wca, Grefa:2022sav}, etc. Recently, machine learning has been applied in the holography~\cite{Akutagawa:2020yeo, Hashimoto:2022eij, Ahn:2024gjf,Ahn:2024jkk,Ahn:2025tjp,Chang:2024ksq,Cai:2024eqa}. A machine learning supplemented EMD framework was developed in Ref.~\cite{Chen:2024ckb}.
In this work, the authors constructed analytical forms of holographic black hole metrics within the EMD framework, applicable to the pure gluon, $2$-flavor and $2+1$-flavor systems, by employing machine learning techniques to fit the equation of state and baryon number susceptibility data from lattice QCD. Meanwhile, this model incorporates the effects of temperature and chemical potential on strongly coupled plasma. Pure Yang-Mills theory ($N_f = 0$) describes the limit without quarks, corresponding to zero quark chemical potential ($\mu_q = 0$). For QCD with $N_f = 2$ (only light $u$ and $d$ quarks), this approximation works well in low-energy regimes where strange quarks are negligible: at low and moderate temperatures. In contrast, $N_f = 2+1$ theory (including dynamical $s$ quarks) is crucial for high-energy experiments like RHIC and LHC, where strange quarks significantly influence strange hadron production ($K^\pm$, $\phi$, $\Lambda$), quark-gluon plasma thermodynamics ($p(T)$, $\epsilon(T)$), and collective flow ($v_2$, $v_3$). Therefore, it can serve as an effective tool for exploring the intricate properties of the strongly coupled QGP and has already been applied to this purpose~\cite{Chen:2024mmd,Chen:2024epd,Guo:2024qiq,Lin:2024mct, Zhu:2025gxo, Li:2025ugv}. However, its limitations-such as Ansatz dependence, sparse training data, and extrapolation uncertainties-highlight the need for further validation (e.g., with more lattice data or alternative machine learning architectures). Future work could explore more flexible Ansatze or unsupervised learning to reduce model bias.

In this paper, we will investigate the instantaneous energy loss of light quarks in the flavor-dependent EMD model assisted by machine learning. We want to understand how the flavor content of the QGP medium through which the jet propagates affects this instantaneous energy loss. On the other hand, we also pay special attention to the contributions of temperature and chemical potential to the light quark energy loss of in strongly coupled QGP, particularly near the phase transition point. The organization of the paper is as follows: In Sec.~\ref{sec:EMD}, the flavor-dependent EMD framework assisted by machine learning, which was proposed in Ref.~\cite {Chen:2024ckb}, will be briefly introduced. In Sec.~\ref{sec:LOSS}, we will provide a short derivation of light quark energy loss formula in the flavor-dependent EMD model by using the shooting string approach. Some numerical results will be presented Sec.~\ref{sec:NUM}, and Sec.~\ref{sec:CON} contains our conclusion and discussion.

\section{the flavor-dependent EMD model }
\label{sec:EMD}
In this section, we review the flavor-dependent EMD background assisted by machine learning~\cite{Chen:2024ckb}. In the Einstein frame,
the following action in five-dimensions can be written as follows:
\begin{equation}\label{e1}
	S = \int \frac{d^{5}x}{16\pi G_{5}} \sqrt{-\text{det}(g_{\mu\nu})} \left[ R - \frac{f(\phi)}{4}F^{2}  - \frac{1}{2}\partial_{\mu}\phi\partial^{\mu}\phi - V(\phi) \right].
\end{equation}
Here $G_5$ is the Newton constant, $F$ is the Maxwell field with field strength tensors $F_{\mu\nu}^{(1)} = \partial_{\mu}A_{\nu} - \partial_{\nu}A_{\mu}$, and $f(\phi)$ corresponds to  the gauge
kinetic function for the Maxwell field. $V(\phi)$ is the potential of the dilaton field. By solving the equations of motion, one can consistently figure out the explicit forms of the gauge kinetic function $f(\phi)$ and the dilaton potential $V(\phi)$.

The metric Ansatz for the metric, gauge field, and dilaton field in the system is assumed by,
\begin{equation}
ds^{2} = \frac{L^{2}e^{2A(z)}}{z^{2}} \left[-g(z)dt^{2} + d\vec{x}^{2} + \frac{dz^{2}}{g(z)} \right],
\end{equation}
where $L$ is the anti¨Cde Sitter (AdS) length scale and $z$ represents the fifth-dimensional holographic coordinate.

The spacetime asymptotic boundary lies at $z = 0$, where
\begin{equation}
A(0)=-\sqrt{\frac{1}{6}} \phi(0), \quad g(0)=1, \quad \phi(0) = 0, \quad A_t(0)=\mu+\rho^{\prime} z^2+\cdots.
\end{equation}
Here, $\mu$ can be interpreted as the baryon chemical potential, which is associated with the quark-number chemical potential through the relation $\mu = 3\mu_q$, and $\rho^{\prime}$ is directly proportional to the baryon number density. Near the horizon of the black hole, it is also subject to the boundary condition that $g(z_t) =A_t(z_t)= 0$, where $z_t$ is the horizon of the black hole, which is related to the temperature.

Following the methodology in \cite{Chen:2024ckb}, the gravity solution can be  given as,
\begin{equation}
\begin{aligned}
g(z)&=1-\frac{1}{\int_0^{z_t} d x x^3 e^{-3 A(x)}}\left[\int_0^z d x x^3 e^{-3 A(x)}+\frac{2 c \mu^2 e^k}{\left(1-e^{-c z_t^2}\right)^2} \operatorname{det} \mathcal{G}\right],\\
\phi^{\prime}(z) & =\sqrt{6\left(A^{\prime 2}-A^{\prime \prime}-2 A^{\prime} / z\right)}, \\
A_t(z) & =\mu \frac{e^{-c z^2}-e^{-c z_t^2}}{1-e^{-c z_t^2}}, \\
V(z) & =-\frac{3 z^2 g e^{-2 A}}{L^2}\left[A^{\prime \prime}+A^{\prime}\left(3 A^{\prime}-\frac{6}{z}+\frac{3 g^{\prime}}{2 g}\right)-\frac{1}{z}\left(-\frac{4}{z}+\frac{3 g^{\prime}}{2 g}\right)+\frac{g^{\prime \prime}}{6 g}\right],
\end{aligned}
\end{equation}
where
\begin{equation}
\operatorname{det} \mathcal{G}=\left|\begin{array}{ll}
\int_0^{z_t} d y y^3 e^{-3 A(y)} & \int_0^{z_t} d y y^3 e^{-3 A(y)-c y^2} \\
\int_{z_t}^z d y y^3 e^{-3 A(y)} & \int_{z_t}^z d y y^3 e^{-3 A(y)-c y^2}
\end{array}\right|.
\end{equation}
The Hawking temperature of this solution is
\begin{eqnarray}
T &=& -\frac{g(z_{t})^{'}}{4\pi}\nonumber\\&=&\frac{z_{t}^{3}e^{-3A(z_{t})}}{4\pi\int_{0}^{z_{t}}dyy^{3}e^{-3A(y)}}[1+\frac{2c\mu^{2}e^{k}(e^{-cz_{t}^{2}}\int_{0}^{z_{t}} dyy^{3}e^{-3A(y)}-\int_{0}^{z_{t}}dyy^{3}e^{-3A(y)-cy^{2}} )}{(1-e^{-cz^{2}_{t}})^{2}}].
\end{eqnarray}
	
To obtain the analytical solutions of this EMD background, one can take the gauge kinetic function $f(z)=e^{c z^2-A(z)+k}$ and the Ansatz of the metric
$A(z)= d \ln(a z^2 + 1) + d \ln(b z^4 +1)$~\cite{Chen:2024ckb}. In Ref.~\cite{Chen:2024ckb}, the authors employed machine learning techniques to construct a six-parameter EMD model applicable to different quark flavor systems by inputting the equation of state (EOS) and baryon number susceptibility data from lattice QCD for the $N_f=0$, $N_f=2$, and $N_f=2+1$ systems, respectively. These  six parameters are shown in Table \ref{1}~\cite{Chen:2024ckb}.
\begin{table}[H]
	\centering
	 \begin{tabular}{|l|c|c|c|c|c|c|}
		\hline
		& $a$ & $b$ & $c$ & $d$ & $k$ & $G_{5}$ \\
		\hline
		$N_{f}=0$ & 0 & 0.072 & 0 & -0.584 & 0 & 1.326 \\
		\hline
		$N_{f}=2$ & 0.067 & 0.023 & -0.377 & -0.382 & 0 & 0.885 \\
		\hline
		$N_{f}=2+1$ & 0.204 & 0.013 & -0.264 & -0.173 & -0.824 & 0.400 \\
		\hline
	\end{tabular}
	\caption{The parameters for the systems with various flavors are used in this model. These data in the table are from Ref.~\cite{Chen:2024ckb}. The unit of $G_5$ is $ \text{GeV}^3$. The units of $a$ and $c$ are $\text{GeV}^2$. The unit of $b$ is $ \text{GeV}^4$.}
\label{1}
\end{table}

For clearer illustration, we present in Fig.~\ref{EOS} a comparative display of the EOS for the systems with different flavors calculated by the machine learning supported EMD model and lattice QCD~\cite{Chen:2024ckb}.

\begin{figure}[H]
\centering
	 \includegraphics[width=0.9\linewidth]{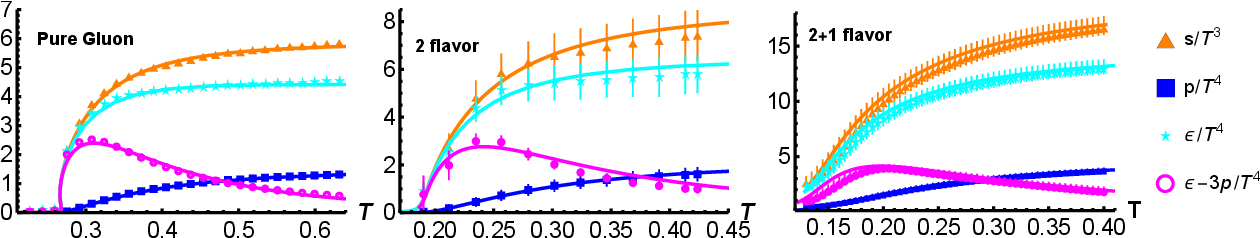}
	\caption{\label{eosall} the EOS with this machine learning supported EMD model and lattice QCD
data for pure gluon, 2-flavor, and 2+1-flavor systems~\cite{Chen:2024ckb}.}
\label{EOS}
\end{figure}

According to the computational methodology in Ref.\cite{Chen:2024ckb}, this machine learning supported EMD model serves as an effective tool for investigating the complex properties of strongly coupled QGP within the holographic framework. If the parameters from Table~\ref{1} are incorporated, this model is able to accurately reproduce the EOS and the baryon number susceptibility observed in lattice QCD for the pure gluon, $2$-flavor, and $2+1$-flavor systems. Furthermore, it includes the temperature and  chemical potential of the critical end points for various flavor systems.
In this paper, based on the analytic form of the holographic black hole metric and relevant parameters in this machine learning augmented EMD framework provided in Ref.[66], we mainly study the instantaneous energy loss of light quarks for the pure gluon, $2$-flavor, and $2+1$-flavor systems, and discuss how the flavor content of strongly coupled plasma affects this instantaneous energy loss. In the next section, we will derive the analytical formula for light quark energy loss within this EMD framework.

\section{ Light quark Energy loss in the flavor-dependent EMD model}
\label{sec:LOSS}
In this section, we will use the holographic model of finite end point momentum shooting string approach proposed in~\cite{Ficnar:2013wba, Ficnar:2013qxa} to study light quark energy loss within the flavor-dependent EMD model. According to Ref.~\cite{Ficnar:2013qxa},  compared with other holographic models of light quark energy loss (e.g., the holographic brick problem ~\cite{Chesler:2014jva, Casalderrey-Solana:2014bpa} and falling strings ~\cite{Chesler:2008wd, Chesler:2008uy}), the shooting string is the simplest phenomenological implementation  of the finite end point momentum method to describe the light quark energy loss in a manner that allows one to calculate observables such as the nuclear modification factor $R_{AA}$ at RHIC and LHC. In the shooting string approach, finite energy and momentum are added to the end points of the shooting strings, enabling it to travel further in the $AdS_5$-Schwarzschild background than the previous falling string configurations~\cite{Chesler:2008uy, Chesler:2008wd}.
Meanwhile, the motion of the shooting string simplifies as follows: the string end point starts near the horizon and then shoots toward the boundary, while the rest of the string sags behind it.
Moreover, the energy loss of light quarks is naturally described as the flow of energy from the end point into the rest of the string during its rise, as the latter embodies the color field generated by the quark. Consequently, it has been proposed that the strings with finite momentum  at their end points may provide a more natural description for energetic light quarks traversing a hot and dense, strongly coupled plasma.

Next, we will provide a short derivation of the light quark energy loss formula in the EMD model by utilizing the shooting string approach. For convenience, we transform to the string frame where $A_{s}(z) = A(z) + \sqrt{1/6}\phi(z)$ in order to calculate the Nambu-Goto action. Then, the metric in the string frame can be rewritten as,
\begin{equation}
\label{metric1}
ds^{2} = \frac{L^{2}e^{2A_s(z)}}{z^{2}} \left[-g(z)dt^{2} + d {x}^{2} + \frac{dz^{2}}{g(z)} \right],
\end{equation}
where we assumed that the end point is moving in the $x$ direction and that the string frame metric $G_{\mu \nu}$ is diagonal in $(t, x, z)$ coordinates and depends only on the fifth-dimensional holographic coordinate $z$, which can be determined according to Eq.(\ref{metric1}) in the EMD background.

From the definition of the end point momenta of the shooting string in Ref.~\cite{Ficnar:2013wba}, we can parametrize $\xi$  and obtain the energy and the momentum (of a test particle) from the metric in Eq.$({\ref{metric1}})$ as follows:
\begin{eqnarray}
\label{pt momentum}
p_t &=&
- \frac{1}{\eta} \frac{L^{2}e^{2A_s(z)}}{z^{2}} g(z) \dot{t} ,
\\
p_x &=&
\frac{1}{\eta} \frac{L^{2}e^{2A_s(z)}}{z^{2}} \dot{x},
\end{eqnarray}
where, as usual, the dot denotes differentiation with respect to some parameter $\xi$ that parametrizes the geodesic with the coordinates $(t, x, z)$, and $\eta$ is an auxiliary field. On the other hand, given that the metric $({\ref{metric1}})$ has no direct relation with $t$, the energy flux flowing from the end point of the string into the rest of the string can be expressed as follows:
\begin{eqnarray}
\label{dotpt}
\dot{p}_t = - \frac{\eta}{2 \pi \alpha' } p_t = - \frac{\sqrt{\lambda}}{2 \pi} \frac{e^{2A_s(z)}}{z^{2}} g(z) \dot{t},
\end{eqnarray}
where $\sqrt{\lambda}= L^2/\alpha'$.
As elaborated in Ref.~\cite{Ficnar:2013wba}, this equation is entirely determined by null geodesics and does not depend on the energy contained within them. Along any geodesic in $AdS_5$, the momenta $p_{t,geo}$ and $p_{x,geo}$ (in the absence of the string) are conserved, and so one can define a purely kinematical quantity $R$ to parametrize the null geodesic. The parameter $R$ is expressed in the following form:
\begin{eqnarray}
\label{parameterR}
R \equiv \frac{p_{t,geo}}{ p_{x,geo}} = - g(z)\frac{\dot{t}}{\dot{x}},
\end{eqnarray}
where $R$ is a constant because the metric $({\ref{metric1}})$ does not depend explicitly on $t$ nor $x$. Consequently, the finite momentum end point will move along null geodesics $d s^2 = 0$. These geodesics could be parametrized by $R$ as follows,
\begin{eqnarray}
\label{null}
\left(\frac{d x_{geo}}{d z}\right)^2
= - \frac{1}{R^2-g(z)}.
\end{eqnarray}
It should be noted that null geodesics in the geometry $({\ref{metric1}})$ are determined by the condition that the denominator of Eq.(\ref{null}) vanishes at $z= z_\ast$. Specifically, $z_\ast$ represents the minimal radial coordinate that the geodesic reaches. Then the geodesic cannot go past that $z_\ast$, and this can be related to $R = - \sqrt{g(z_\ast)}$.

Using the $\xi =x$ parametrization and substituting Eq.(\ref{parameterR}) into Eq.(\ref{dotpt}), one can obtain the instantaneous energy loss as
\begin{eqnarray}
\label{energyloss}
\frac{d E}{d x}  = - \frac{\sqrt{\lambda}}{2 \pi } \frac{|R| e^{2A_s(z)}}{z^{2}}
.
\end{eqnarray}
Here we identify $p_t$ with $-E$ at the boundary and for the case where the end point energy decreases with time, the ``$-$'' sign is selected. The above Eq.(\ref{energyloss})  represents  the analytical expression for the instantaneous energy loss of light quarks within the flavor-dependent EMD model. To better understand the nature of the light quark energy loss, we will perform a numerical integration of Eq.(\ref{null}) in the small $z_\ast$ limit and invert it to obtain $z(x)$ . Then, it can be substituted into Eq.(\ref{energyloss}) to get $d E/d x$ as a function of $x$. It should be noted that in the  $z_\ast$ limit for asymptotically $AdS_5$ geometries, one usually sets $z_\ast\rightarrow 0$ (and $R\rightarrow1$ ) in the shooting string approach~\cite{Ficnar:2013wba, Ficnar:2013qxa, Rougemont:2015wca, Zhu:2019ujc, Zhang:2019jfq, Zhang:2019gki, Zhu:2020wds, Zhang:2023kzf}.

Before proceeding to discuss  these results, we need to recall the supersymmetric Yang-Mills (SYM)results.SYM results. If one utilizes the limit of $z<<z_t$ and only considers the contribution of the leading order in the Taylor expansion of the metric Ansatz function $A(z)$ in the systems with various flavors within the EMD model without chemical potential $\mu$, we have the capacity to obtain the subsequent approximate analytical gravity solutions for the EMD background,
\begin{eqnarray}
A_s(z)\approx1,\;\;\; g(z)\approx 1-\frac{z^4}{z_t^4},\;\;\; T \approx \frac{1}{\pi z_t}.
\end{eqnarray}
Then, we can solve the null geodesic equation (\ref{null})  to obtain the approximate analytical solution of $x_{geo}(z)$ as,
\begin{eqnarray}
x_{geo}(z)\approx z_t^2\left(\frac{1}{z}-\frac{1}{z_0}\right),
\end{eqnarray}
where we assume that the string is initially a point at some radial coordinate $z_0$. With the above simplifications, Eq.(\ref{energyloss}) can be reduced to the following expression for the light quark energy loss in $\mathcal{N} = 4$ SYM plasma~\cite{Ficnar:2013wba, Ficnar:2013qxa},
\begin{eqnarray}
\label{SYMenergyloss}
\left(\frac{d E}{d x}\right)_{SYM} = - \frac{\pi \sqrt{\lambda}}{2 } T^2 (\frac{1}{\tilde{z}_0} + \pi T x),
\end{eqnarray}
where $\tilde{z}_0 \equiv \pi T z_0 \in [0, 1]$. In this paper, we choose $\tilde{z}_0 = 1$, similar to Ref.~\cite{Ficnar:2013qxa}.

At this stage, we have provided the crucial analytical formula, seeing Eqs. (\ref{null}) and (\ref{energyloss}), for the light quark energy loss formula in the EMD model by applying the shooting string approach. In the next section, we will discuss some numerical results of instantaneous energy loss of light quarks to analyze its properties.

\section{numerical results}
\label{sec:NUM}
In this section, we intend to put forward some numerical results based on the analytical calculations mentioned in the previous section.
While the analytical form for light quark energy loss [see Eq.(\ref{energyloss})]  is consistent across different flavor systems, we can calculate the light quark energy loss for various flavor systems individually by incorporating the six parameters listed in tab Table~\ref{1}, which are provided by Ref.~\cite{Chen:2024ckb} using machine learning techniques, for the pure gluon, 2-flavor, and 2+1-flavor systems, specifically.
Now, we first aim to explore the dependence of the light quark energy loss on the flavor content of strongly coupled plasma. In Fig.~\ref{fig2}, we compare the instantaneous energy loss of the EMD model at finite temperature and zero chemical potential to that of the SYM system, where Fig.~\ref{fig2a} is for $T = 0.3 \,\text{GeV}$ and Fig.~\ref{fig:subfig:2b} is for $T = 0.5 \,\text{GeV}$.
We can find that this ratio decreases as the value of $x$ increases at a fixed temperature. Moreover, the results indicate that $(d E/d x)/(d E/d x)_{\text{SYM}}$ is smallest for $N_f =0$, slightly larger for $N_f = 2$, and the greatest for $N_f = 2+1$. Namely, the addition of $u (d)$ quark leads $(d E/d x)/(d E/d x)_{\text{SYM}}$ to increase, and the addition of the $s$ quark further enhances this ratio.
\begin{figure}[H]
\centering
\subfigure[]{
\label{fig2a}
\includegraphics[width=0.45\linewidth]{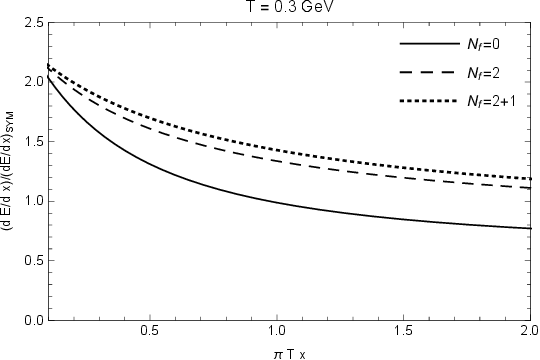}}
\subfigure[]{
\label{fig:subfig:2b}
\includegraphics[width=0.45\linewidth]{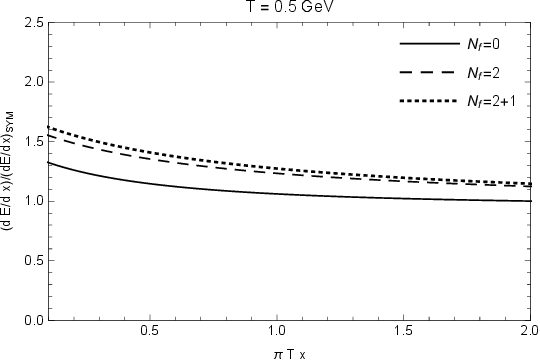}}
\caption{ Ratio between the instantaneous energy loss in the EMD model without chemical potential [seeing Eq.(\ref{energyloss})] and that in the SYM model [seeing Eq.(\ref{SYMenergyloss})] as a function of $\pi\, T\,x$ for different flavor systems: (a) with $T = 0.3 \,\text{GeV} $, (b) with $T = 0.5 \,\text{GeV} $. }
\label{fig2}
\end{figure}

Next, we continue to study the influence of temperature on the light quark energy loss in the flavor-dependent EMD background at vanishing chemical potential. We plot $(d E/dx)/(-\sqrt{\lambda})$ as a function of $T$ for different values of $x$ in Fig.~\ref{fig3}. Specifically, Fig.~\ref{fig3a} corresponds to the pure gluon system, Fig.~\ref{fig:subfig:3b} to the $2$-flavor system, and Fig.~\ref{fig:subfig:3c} to $2+1$-flavor system. In our work, we only consider the energy loss in the deconfined state, where the critical temperature of the various flavor-dependent systems at vanishing chemical potential was predicted in Ref.~\cite{Chen:2024ckb}: $T_c = 0.265 \,\text{GeV}$ for the pure gluon system, $T_c = 0.189\,\text{GeV}$ for the $2$-flavor system, and $T_c = 0.128\,\text{GeV}$ for the $2+1$-flavor system. It can be observed that, in various flavor systems, the instantaneous energy loss of light quarks is minimal near the critical temperature $T_c$ and gradually increases as the temperature $T$ rises. Moreover, the larger the value of shooting distance $x$, the more significant this increasing trend becomes. On the other hand, through longitudinal comparison, we find that the energy loss at a fixed temperature value increases with the growth of $x$.
\begin{figure}[H]
\centering
\subfigure[]{
\label{fig3a}
\includegraphics[width=0.3\linewidth]{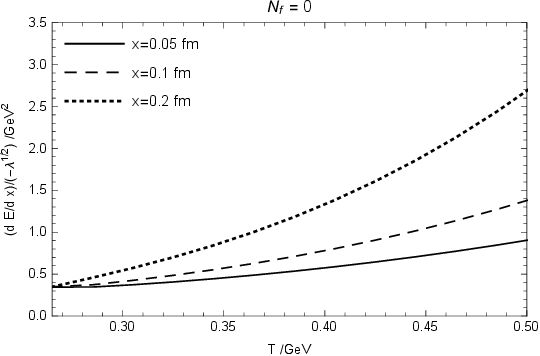}}
\subfigure[]{
\label{fig:subfig:3b}
\includegraphics[width=0.3\linewidth]{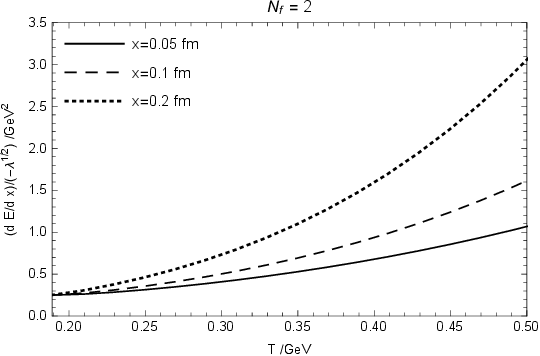}}
\subfigure[]{
\label{fig:subfig:3c}
\includegraphics[width=0.3\linewidth]{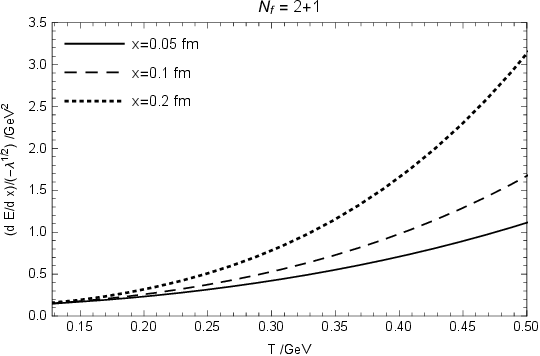}}
\caption{$(d E/dx)/(-\sqrt{\lambda})$ as a function of $T$ with zero chemical potential for different values of $x$, corresponding to (a)  $N_f = 0$, (b) $N_f = 2$, (c) $N_f = 2+1$. }
\label{fig3}
\end{figure}

Based on the parameters in Table~\ref{1} \cite{Chen:2024ckb}, it can be  found that the pure gluon system is independent of the chemical potential, whereas the metrics of the $N_f = 2$ and $N_f = 2+1$ systems explicitly depend on it. Therefore, we also focus on the influence of the chemical potential on the instantaneous energy loss in the $2$-flavor and $2+1$-flavor systems. In Fig.~\ref{fig4}, we plot $(dE/dx)/(-\sqrt{\lambda})$ against $\pi\, T\, x$ at a fixed temperature $T$ and different values of chemical potential $\mu$ for the $2$-flavor system and the $2+1$-flavor systems, respectively. To better analyze the energy loss of light quarks near the critical end point in the strongly coupled QGP, we set $T = 0.15 \,\text{GeV} $ in Fig.~\ref{fig4a} and $T = 0.1 \,\text{GeV} $ in Fig.~\ref{fig:subfig:4b}, which represent relatively low temperatures close to the critical temperatures of the $2$-flavor and $2+1$-flavor systems, respectively. According to Ref.~\cite{Chen:2024ckb}, the predicted critical end point for the $2$-flavor system is located at $T_c = 0.147 \,\text{GeV}, \mu_B^c = 0.46 \,\text{GeV}$ and for the $2+1$-flavor system is at $T_c = 0.094 \,\text{GeV}, \mu_B^c = 0.74 \,\text{GeV}$. As depicted in these figures,
when the chemical potential in Fig.~\ref{fig4a} and Fig.~\ref{fig:subfig:4b} is respectively set to $0.5$ and $0.8 \,\text{GeV}$ (see the solid lines in these figures), the light quark energy loss in both flavor-dependent systems is minimal and exhibits an almost negligible increase as $x$ grows. Namely, this energy loss near the critical end point remains consistently small, which is consistent with the results shown in Fig.~\ref{fig3} for the case at vanishing chemical potential. However, when we take a larger chemical potential, the system moves away from the critical endpoint, and the instantaneous energy loss varies significantly with $x$.
Furthermore, in Figs.~\ref{fig6} and ~\ref{fig7}, we respectively present the curves of the energy loss of light quarks in both flavor-dependent systems as a function of $\mu$ when $T$ and $x$ are fixed. It indicates that as the temperature and chemical potential increase, the system gradually deviates from the critical end point, and the light quark energy loss will rise. The physical significance of these results will be discussed in the following section.

\begin{figure}[H]
\centering
\subfigure[]{
\label{fig4a}
\includegraphics[width=0.4\linewidth]{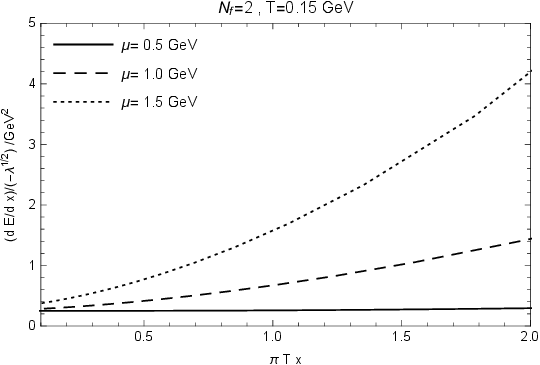}}
\subfigure[]{
\label{fig:subfig:4b}
\includegraphics[width=0.4\linewidth]{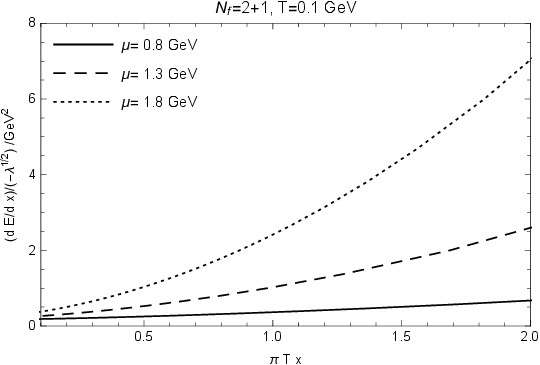}}
\caption{$(d E/dx)/(-\sqrt{\lambda})$ versus $\pi\, T\, x$ for a fixed $T$ and different values of $\mu$. (a)for the $N_f=2+1$ system, $T = 0.15 \,\text{GeV}$ and $\mu= 0.5,\, 1,\, 1.5 \,\text{GeV}$ for the $N_f=2$ system ; (b) for the $N_f=2+1$ system, $T = 0.1 \,\text{GeV}$, $\mu= 0.8,\, 1.3,\, 1.8 \,\text{GeV}$.}
\label{fig4}
\end{figure}

\begin{figure}[H]
\centering{
\includegraphics[width=0.5\linewidth]{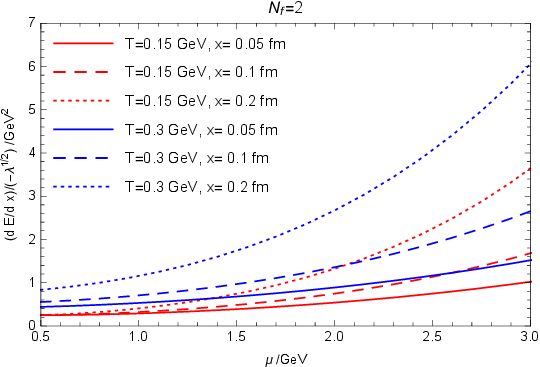}}
\caption{The curves of the energy loss in $2$-flavor system as a function of $\mu$ with fixed $x$ and $T$. Here the red curves are corresponding to $T = 0.15 \,\text{GeV} $, and the blue curves are corresponding to $T = 0.3 \,\text{GeV} $. The solid, dashed and dotted lines correspond to $x = 0.05$, $0.1$, and $0.2\,\text{fm}$, respectively.}
\label{fig6}
\end{figure}

\begin{figure}[H]
\centering{
\includegraphics[width=0.5\linewidth]{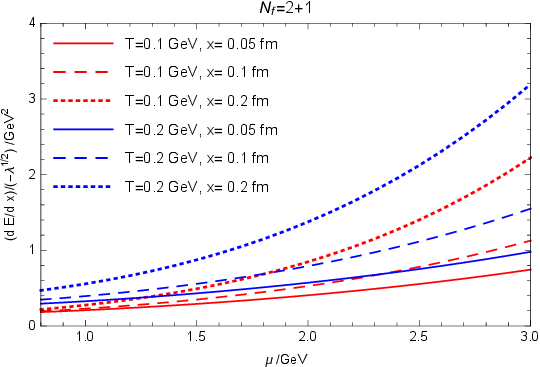}}
\caption{The curves of the energy loss in the $2+1$-flavor system as a function of $\mu$ with fixed $x$ and $T$. Here the red curves are corresponding to $T = 0.1 \,\text{GeV} $, and the blue curves are corresponding to $T = 0.2 \,\text{GeV} $. The solid, dashed and dotted line  correspond to $x = 0.05$, $0.1$, and $0.2\,\text{fm}$, respectively.  }
\label{fig7}
\end{figure}

\section{Conclusion and discussion}
\label{sec:CON}
The phenomenon of jet quenching is crucial for studying the properties of QGP that was generated in high-energy heavy-ion collisions. Studying jet quenching in strongly coupled plasma at finite temperature and chemical potential can provide valuable insights into the properties of QGP. In this paper, by utilizing the shooting string approach, we investigated the light quark energy loss in the flavor-dependent EMD model. Specifically, we discussed how the flavor content of the strongly coupled plasma, temperature, chemical potential, and the shooting distance of the string end point of the shooting string affect this energy loss. First, we calculated the instantaneous energy loss of light quarks in various flavor-dependent systems without chemical potential. Our results showed that the addition of either $u(d)$ quarks or $s$ quark increases the ratio of the energy loss in the EMD model and that in the SYM system. Furthermore, when the temperature is close to the critical temperature, the energy loss of light quarks in the system is minimal, but it increases significantly as the temperature gradually rises.
Second, we separately analyzed the instantaneous  energy loss of light quarks in the $2$-flavor and the $2+1$-flavor systems at finite temperature and finite chemical potential. We found that the light quark energy loss gradually increases as the system moves away from the critical end point with increasing temperature or chemical potential. These results are consistent with those derived from the drag force and jet quenching parameter within the same model~\cite{Chen:2024epd}.

As is well known, instantaneous energy loss is indeed crucial in phenomenological jet quenching applications. For example, both the calculation of the nuclear modification factor $R_{AA}$ and the elliptic flow parameter $v_2$ of light hadrons require detailed knowledge of the light quark energy loss.
In future efforts, we will apply our results on light quark energy loss to phenomenological jet quenching studies in strongly coupled plasma created in the high-energy nuclear collisions.

\section*{ACKNOWLEDGMENTS}

This work is supported by the National Natural Science Foundation of China (NSFC) under Grants No. 12405154 and No. 12005056, and Hubei Provincial Natural Science Foundation Youth Project under Grant No. 2024AFB151.

\end{widetext}
%

\bibliographystyle{plain}
\bibliographystyle{h-physrev5}
\bibliography{zladscft_refs}
\end{document}